\documentclass[aps, 10pt, prb, amssymb, eqsecnum, twocolumn, floatfix, superscriptaddress]{revtex4-2}
\usepackage[mathscr]{euscript}
\usepackage{bm}
\usepackage{euscript}
\usepackage{exscale}
\usepackage{amsbsy}
\usepackage{textcomp}
\usepackage{comment}
\usepackage{slashed}
\usepackage{tabularx}
\usepackage{euscript}
\usepackage{exscale}
\usepackage{wrapfig}
\usepackage{makecell}
\usepackage{amsmath,amsfonts,amsthm} 
\usepackage{mathtools}
\usepackage{graphicx} 
\usepackage{bbold} 
\usepackage{physics}
\usepackage{hyperref} 
\usepackage{xcolor} 
\usepackage{enumitem} 
\usepackage{caption}
\usepackage{subcaption}
\usepackage{float}
\newcommand{\id}{\mathbb{1}}

\def\sgn{\text{sgn}}
\def\hc{\text{h.c.}}

\newcommand{\pbr}[1]{\left(#1\right)}
\newcommand{\sbr}[1]{\left[#1\right]}
\newcommand{\cbr}[1]{\left\{#1\right\}}

\def\Z{\mathbb{Z}}

\def\L{\mathcal{L}}

\def\bk{\braket}

\def\da{\dagger}

\def\rd{{\rm d}}
\def\pa{\partial}

\def\ra{\rightarrow}

\def\al{\alpha}
\def\be{\beta}
\def\de{\delta}
\def\De{\Delta}

\def\Ga{\Gamma}
\def\la{\lambda}

\def\om{\omega}

\def\pa{\partial}

\def\vp{\varphi}
\def\vt{\vartheta}
\def\si{\sigma}

\def\Th{\Theta}

\hypersetup{
    unicode=false,          
    pdftoolbar=true,        
    pdfmenubar=true,        
    pdffitwindow=false,     
    pdfstartview={FitH},    
    pdftitle={PZM in FTSC},    
    pdfauthor={},     
    pdfsubject={Subject},   
    pdfcreator={},   
    pdfproducer={}, 
    pdfkeywords={keyword1} {key2} {key3}, 
    pdfnewwindow=true,      
    colorlinks=true,       
    linkcolor=blue, 
    citecolor=blue,        
    filecolor=magneta,      
    urlcolor=cyan           
} 
\begin{document}

\title{Signatures of Parafermion Zero Modes in Fractional Quantum Hall-Superconductor Heterostructures}
\author{Junyi Cao}
\affiliation{Department of Physics, University of Illinois Urbana-Champaign, 1110 West Green Street, Urbana, Illinois 61801, USA}
\affiliation{The Anthony J. Leggett Institute for Condensed Matter Theory, University of Illinois Urbana-Champaign, 1110 West Green Street, Urbana, Illinois 61801, USA}
\author{Angela Kou}
\affiliation{Department of Physics, University of Illinois Urbana-Champaign, 1110 West Green Street, Urbana, Illinois 61801, USA}
\affiliation{Materials Research Laboratory, University of Illinois Urbana-Champaign, 104 S. Goodwin Ave, Urbana, IL 61801, USA}
\author{Eduardo Fradkin}
\affiliation{Department of Physics, University of Illinois Urbana-Champaign, 1110 West Green Street, Urbana, Illinois 61801, USA}
\affiliation{The Anthony J. Leggett Institute for Condensed Matter Theory, University of Illinois Urbana-Champaign, 1110 West Green Street, Urbana, Illinois 61801, USA}

\begin{abstract}
	Parafermion zero modes can arise in hybrid structures composed of $\nu=1/m$ fractional quantum Hall edges proximity coupled by an s-wave superconductor. Here we consider a Josephson junction formed in such hybrid structures in addition to parafermion tunneling, Cooper pair tunneling, and backscattering. We find that the $4\pi m$ periodicity due to parafermion-only tunneling reduces, in the presence of backscattering, to $4\pi$-periodic at zero temperature and $2\pi$-periodic at finite temperature unless the fermion parity is fixed. Nevertheless, a clear signature of parafermion tunneling remains in the shape of the current-phase relation.
\end{abstract}
\maketitle
\section{Introduction}
Non-Abelian topologically ordered phases are among the most promising platforms for fault-tolerant quantum computation \cite{nayak-2008}. The excitations in these phases are non-Abelian anyons that have non-trivial fusion rules and braiding statistics \cite{book-2013}. These fusion rules provide a source of topological ground state degeneracy, which allows for non-local storage of information and anyon braiding that is topologically protected against decoherence \cite{chamon-1997,fradkin-1998,bonderson-2006,stern-2006,bishara-2009}. While there has been significant interest in using Majorana zero modes (MZMs) for topological quantum computation \cite{nayak-2008}, the $\Z_N$ generalization of MZMs, parafermion zero modes (PZMs) \cite{fradkin-1980}, is necessary to perform universal topological quantum computation. It was shown in Ref. \cite{mong-2014} that an array of PZMs provides a realization of Fibonacci anyons that is capable of universal topological quantum computation \cite{nayak-2008,freedman-2000,hastings-2013,hastings-2014}.

It has been theoretically proposed that PZMs can arise in fractional topological superconductors (FTSCs) \cite{clarke-2013,lindner-2012,cheng-2012}. A key example of an FTSC comprises edge states of a $\nu=1/m$ fractional quantum Hall (FQH) system proximitized with an s-wave superconductor. The physics of proximitized FQH edges is particularly relevant at the moment due to recent experiments demonstrating the viability of experimental setups to manipulate and control parafermions in hybrid structures at moderate magnetic fields \cite{du-2009,dean-2011}. In particular, a recent experiment has focused on implementing such a structure in graphene and observed crossed Andreev reflection (CAR), which was suggested to indicate the presence of PZMs \cite{gul-2022}. A theoretical analysis in Ref. \cite{schiller-2022} showed that CAR is a necessary but not sufficient condition for the existence of PZMs. 

In this letter, we consider an FTSC resulting from the proximity effect between an s-wave superconductor and two edges of a $\nu=1/m$ FQH state and identify the PZM at one end. We demonstrate that a Josephson junction consisting of two copies of such FTSCs captures unique features due to the fractionalization of PZMs. In the low-energy effective Hamiltonian, we determine the energy spectra and current-phase relation in the presence of parafermion tunneling, Cooper pair tunneling, and backscattering. We find that backscattering explicitly breaks the $\Z_m$ symmetry present in the junction, which results in the periodicity of the Josephson phase being the same for PZMs and MZMs. While the periodicity is an insufficient distinguishing metric, additional features arise in the thermally-averaged current-phase relation that discriminates between PZMs and MZMs. As an alternative measure, we propose the parity-projected thermally-averaged current-phase relation which results in a temperature-dependent $4\pi$-periodic fractional Josephson effect that occurs only under the presence of parafermion tunneling.

\section{Model for FTSC and PZM}
\label{sec:model}
\begin{figure}
    \centering
    \includegraphics[width=.9\linewidth]{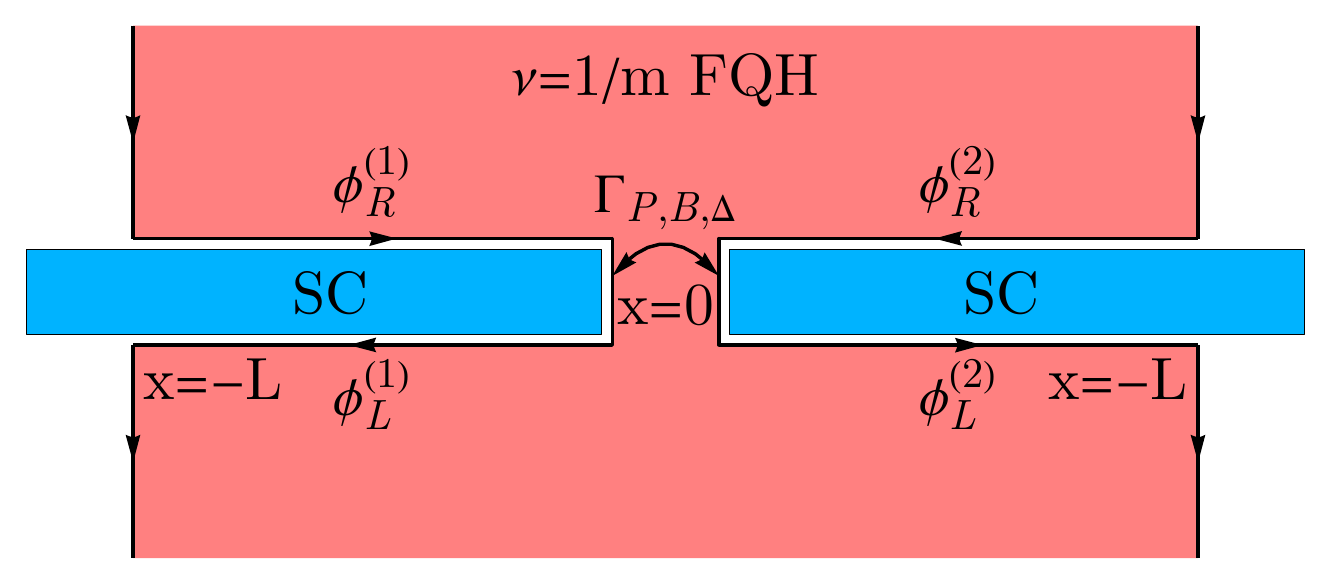}
    \caption{The Josephson junction includes two copies of FTSCs, parafermion tunneling $\Ga_P$, Cooper pair tunneling $\Ga_\De$, and backscattering $\Ga_B$. Each FTSC consists of two edges from $\nu=1/m$ FQH system (pink) proximity coupled by an s-wave SC (cyan). The spacing between the two FTSCs is exaggerated to emphasize the presence of a junction.}
    \label{fig:FTSCjunc}
\end{figure}

We first discuss our model for an FTSC, which consists of two edges from a (fully spin-polarized) $\nu=1/m$ FQH system proximity coupled by an s-wave superconductor (SC) finger as in Fig. \ref{fig:FTSCjunc}. Since at each finger, the edges are in proximity with an s-wave superconductor finger, we include a density-density interaction term and a pairing term $\De\psi_L\psi_R+\hc$ Here we denote by $L$ and $R$ the top and bottom FQH edges states in contact with the SC finger. The corresponding bosonized (see Appendix \ref{sec:bosonization} for reviews on bosonization) Hamiltonian is,
\begin{align}
\begin{split}
    H=&\int_{-L}^0 \rd x~\bigg\{\frac{mv_F}{4\pi}\left[\left(\pa_x\phi_R\right)^2+\left(\pa_x\phi_L\right)^2\right]\\
    &+\frac{mU}{2\pi}\pa_x\phi_R\pa_x\phi_L-\frac{{\De}}{\ell_0^2}\cos\left[m\left(\phi_R-\phi_L\right)\right]\bigg\},
\end{split}
\end{align}
where $v_F$ is the Fermi velocity of the edges, $\phi_{R/L}$ are the bosonic fields on the top/bottom edge, $U$ is the interaction strength, ${\De}>0$ is the (dimensionless) proximity gap, and $\ell_0$ is the magnetic length which plays the role of a UV cutoff.

It is useful to rewrite the bosonic fields as $\phi_{R/L}=\vp\pm\vt$, which obey the commutation relation,
\begin{align}
	\comm{\vp(x)}{\vt(x')}=i\frac{\pi}{m}\Th(x-x'),\label{eqn:comm}
\end{align}
where $\rho\equiv\pa_x\vp/\pi=\rho_{R}+\rho_{L}$ is the total charge density operator. The Hamiltonian in the new basis is,
\begin{align}
\begin{split}
	H=\int_{-L}^0 \rd x\bigg\{&\frac{mu}{2\pi}\left[K\left(\pa_x\vp\right)^2+\frac{1}{K}\left(\pa_x\vt\right)^2\right]\\
	&-\frac{{\De}}{\ell_0^2}\cos(2m\vt)\bigg\},
\end{split}
	\label{eqn:H}
\end{align}
with effective velocity $mu/\pi$ with $u=\sqrt{v_F^2-U^2}$ which we set to 1 in the following sections and Luttinger parameter $K=(v_F+U)/\sqrt{v_F^2-U^2}$, making this theory a sine-Gordon theory on a finite interval. In the vicinity of the critical point, the proximity gap $\De$ and the distance of the Luttinger parameter to the critical point $x\equiv 2-mK$ parametrize Kosterlitz RG \cite{kosterlitz-1974,Amit-1980} (see Appendix \ref{sec:RG} for details of RG analysis),
\begin{align}
	\dv{{\De}}{l}=x {\De},~\quad \dv{x}{l}=128m^2\pi^5{\De}^2.
\end{align}
For the FTSC to be in the superconducting phase, either Luttinger parameter must satisfy $K<K_c=2/m$ \footnote{The Luttinger parameter we use here is related to the Luttinger parameter in \cite{schiller-2022} by $K\ra 1/K$.} for the superconducting term to be relevant, which requires a sufficiently large attractive interaction, or strong pairing ${\De}>{\De}_c$ is required. This is different from the $\nu=1$ case where the superconducting term is always relevant even in the absence of a density-density interaction ($K=1$) (see Fig. \ref{fig:RG}). Deep in the superconducting phase, the $\vt$ field is pinned to one of the $2m$ minima of the cosine term in Eq. \eqref{eqn:H}, $\vt(x)=\frac{\Tilde{n}\pi}{m}$, where $\Tilde{n}\in\Z_{2m}$ is an integer-valued operator which is related to the clock operator in the $N$-state clock model on a single site with $N=2m$ \cite{book-2013} (see Appendix \ref{sec:clock} for review on $N$-state clock model). 

This Hamiltonian has a $\Z_{2m}$ symmetry, $\vt\ra\vt+\frac{\pi}{m}$, representing charge conservation modulo $e/m$. The corresponding symmetry generator is a $\Z_{2m}$ generalization of the fermion parity operator $(-1)^{F}$, which we call quasiparticle parity,
\begin{align}
	\hat{P}\equiv \exp(i\pi \hat{Q})=\exp\sbr{i\pbr{\vp_0-\vp_{-L}}},
\end{align}
where $Q\equiv\int_{-L}^0\rd x~\rho(x)$ is the number operator and $\vp_{0}\equiv\vp(x=0)$. One can check that this is indeed the symmetry generator since,
\begin{align}
	\hat{P}^{-1} \vt \hat{P} =\vt+\frac{\pi}{m}.
\end{align}
In the ground state, the $\Z_{2m}$ symmetry sends the system from one pinned minima to another $\tilde{n}\ra\tilde{n}+1\mod 2m$, implying a $2m$-fold ground state degeneracy and the existence of parafermion zero modes at the ends of the FTSC (see Appendix \ref{sec:clock} for review on PZM algebra). The parafermion operator localized at one end of the FTSC is,
\begin{align}
	\al_0=\frac{1}{b}\int_{-b}^0 \rd x~e^{i\vt(x)}\propto e^{i\vt_0},
\end{align}
where $\vt_0\equiv\vt(x=0)$, and $b$ denotes the length of the region where a PZM is localized and is comparable to the coherence length $\xi$. From the solutions of the sine-Gordon equation, $\vt$ has exponentially small fluctuations for $x\in[-b,0]$; hence, we can treat $e^{i\vt(x)}$ as a constant in this region as well. From the form of the PZM, we can interpret it as ``half" of a quasiparticle pair, which reflects the fractionalized nature of the system. One can check $\al_0$ is the PZM operator by the commutator between the Hamiltonian and the PZM,
\begin{align}
	\comm{H}{\al_0}&=-K\pa_x\vp_0e^{i\vt_0}=0.
\end{align}
The commutator vanishes since the total current density $j=\pa_x \vp/\pi$ is 0 at $x=0$, representing no total current flowing from the edge to the FQH background \footnote{This boundary condition is equivalent to treating the FQH bulk (where the bosonic fields are not well defined due to absence of edges) as FQH edges gapped by backscattering term $\cos(2m\vp)$ as described in Ref. \cite{clarke-2013,lindner-2012,cheng-2012}. In the regime where the backscattering term dominates, the $\vp$ field is pinned to one of the $2m$ minima of the cosine term, implying $\pa_x\vp_0=0$.}. The commutation relation between the parafermion zero mode operator $\al_0$ and the quasiparticle parity $\hat{P}$ is,
\begin{align}
	\hat{P}\al_0&=e^{i\frac{\pi}{m}} \al_0\hat{P}.
\end{align} 
Physically this means that the parafermion operator changes the number operator by one. 

\section{Josephson junction and tunneling}
\label{sec:junction}
One way to experimentally identify topologically nontrivial zero modes is the fractional Josephson effect \cite{kitaev-2001,clarke-2013,lindner-2012,cheng-2012}, i.e. $4m\pi$-periodic signal in the current-phase relation. The Josephson junction of interest consists of two copies of FTSCs as in Fig. \ref{fig:FTSCjunc},
\begin{align}
    \begin{split}  
    H=\int_{-L}^0 &\rd x~\bigg\{\sum_{i=1,2}\frac{m}{2\pi}\left[K\left(\pa_x\vp^{(i)}\right)^2+\frac{1}{K}\left(\pa_x\vt^{(i)}\right)^2\right]\\
    &-\frac{\De}{\ell_0^2}\cos(2m\vt^{(1)})-\frac{\De}{\ell_0^2}\cos(2m\vt^{(2)}-\de\phi_{sc})\bigg\},
    \end{split}
    \label{eqn:hjunc}
\end{align}
where $\de\phi_{sc}$ is the Josephson phase between the two SCs, and $i=1,2$ describes the edges near the left/right SC, respectively. This system has symmetries including an overall $\Z_{2m}$ quasiparticle parity and charge conjugation $\vt^{(i)}\ra-\vt^{(i)},~\vp^{(i)}\ra\vp^{(i)}$. The allowed tunneling between two FTSCs are parafermion tunneling (charge $e/m$), Cooper pair tunneling (charge $2e$), and backscattering (charge 0). There are additional symmetry-allowed terms involving tunneling of a group of parafermions, e.g. Majorana fermion tunneling can be understood as tunneling of a group of $m$ parafermions. Here we study the simplest terms as higher-order terms are exponentially suppressed.

The parafermion tunneling Hamiltonian can be written as,
\begin{align}
\begin{split}
H_P&=\Ga_P \al^{(1)\da}_0\al^{(2)}_0+\hc\\
&=\Ga_P e^{i\pbr{\vt^{(1)}_0-\vt^{(2)}_0-\frac{\de\phi_{sc}}{2m}}}+\hc,
\end{split}
\end{align}
where $\Ga_P$ is the parafermion tunneling amplitude and for $\nu=1$, $\Ga_P$ represents Majorana tunneling. For clarity, we will denote the Majorana tunneling amplitude as $\Ga_M$. Similarly, the Cooper pair tunneling can be written as,
\begin{align}
    H_\De&=\Ga_\De e^{2mi\pbr{\vt^{(1)}_0-\vt^{(2)}_0-\frac{\de\phi_{sc}}{2m}}}+\hc
\end{align}
We can see the Josephson phase $\de\phi_{\text{sc}}$ is $4m\pi$-periodic.

Another process consistent with symmetry is the tunneling of ``half" of a quasiparticle-quasihole pair, which we call backscattering,
\begin{align}
H_B&=\Ga_B e^{i\pbr{\vp^{(1)}_0-\vp^{(2)}_0}}+\hc
\end{align}
This tunneling term tunnels 0 charge; hence, it does not couple to the electromagnetic field and does not contribute to tunneling current.

To see the behaviors of these tunneling terms, we consider energies below the superconducting gap, $\abs{E}\ll \De$ \footnote{At higher energies, the inclusion of the bulk in Eq. \eqref{eqn:hjunc} changes the correlation functions of the bosonic fields and response functions, but signatures of PZMs we discuss later will not be affected.}, where the system can be described by an effective Hamiltonian including all of the tunneling processes. The effective Hamiltonian can then be written in the basis where $\exp(i\vp^{(i)}_0)$ are diagonal, with eigenvalues $\exp(in^{(i)}\pi/m)$, where $n^{(i)}$ can be thought as the eigenvalue of the number operator $n^{(i)}\in\Z_{2m}$. In this basis, the parafermion operator $e^{i\vt^{(i)}_0}$ shifts $n^{(i)}$ by one and the Cooper pair tunneling term shifts $n^{(i)}$ by $2m$, which is equivalent to not changing $n^{(i)}$. The effective Hamiltonian is,
\begin{widetext}
\begin{align}
\begin{split}
	H_{\text{eff}}=\sum_{n^{(1)},n^{(2)}=0}^{2m-1} &\cbr{2\abs{\Ga_B}\cos\sbr{\frac{(n^{(1)}-n^{(2)})\pi}{m}}+2\abs{\Ga_\De}\cos(\de\phi_{sc})}\ketbra{n^{(1)},n^{(2)}}\\
    &+\abs{\Ga_P}\left(e^{-i\frac{\de\phi_{sc}}{2m}}\ketbra{n^{(1)}+1,n^{(2)}-1}{n^{(1)},n^{(2)}}+e^{i\frac{\de\phi_{sc}}{2m}}\ketbra{n^{(1)},n^{(2)}}{n^{(1)}+1,n^{(2)}-1}\right).
\label{eqn:heff}
\end{split}
\end{align}
\end{widetext}
Since $n^{(i)}\in\Z_{2m}$ and all tunneling terms conserve the total quasiparticle parity $n^{(1)}+n^{(2)}\mod2m$, this Hamiltonian can be block diagonalized. In the following, we only consider the effective Hamiltonian in the sector where the total quasiparticle parity is zero. Sectors with nonzero quasiparticle parity have spectra that differ by multiples of $2\pi$. The wavefunction $\Psi_r(\de\phi_{sc})$ satisfies a Harper-like equation \cite{wen-1989},
\begin{align}
\begin{split}
    &\abs{\Ga_P}\sbr{e^{-i\frac{\de\phi_{sc}}{2m}}\Psi_{r+1}(\de\phi_{sc})+e^{i\frac{\de\phi_{sc}}{2m}}\Psi_{r-1}\pbr{\de\phi_{sc}}}\\
    &+\sbr{2\abs{\Ga_B}\cos(2\pi \frac{r}{m})+2\abs{\Ga_\De}\cos(\de\phi_{sc})}\Psi_r(\de\phi_{sc})\\
    &=E_r(\de\phi_{sc})\Psi_r\pbr{\de\phi_{sc}},
\end{split}
\end{align}
with $1\leq r\leq 2m$. The eigenstates satisfy the periodic boundary condition $\Psi_r\pbr{\frac{\de\phi_{sc}}{2m}+2\pi k}=\Psi_r\pbr{\frac{\de\phi_{sc}}{2m}}$ with $k\in\Z$.

\begin{figure}[!ht]
\centering
    \begin{subfigure}{.5\linewidth}
         \centering
         \hspace*{-0.45cm}\includegraphics[width=\linewidth]{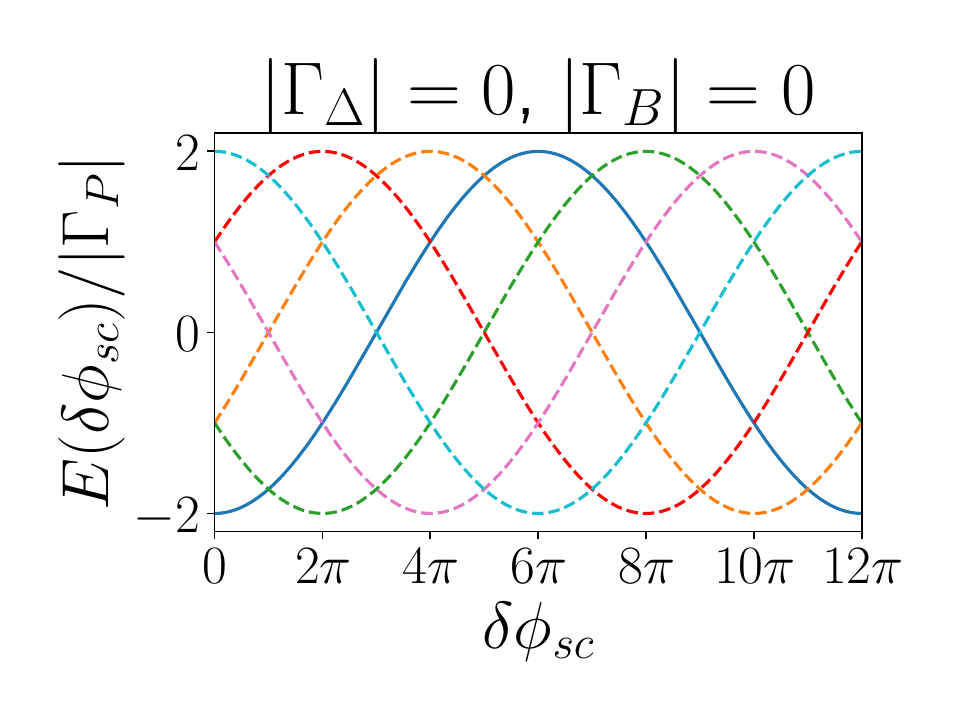}
         \captionsetup{justification=centering}
         \caption{}
         \label{fig:heff_P}
    \end{subfigure}\hfill
    \begin{subfigure}{.5\linewidth}
         \centering
         \hspace*{-0.45cm}\includegraphics[width=\linewidth]{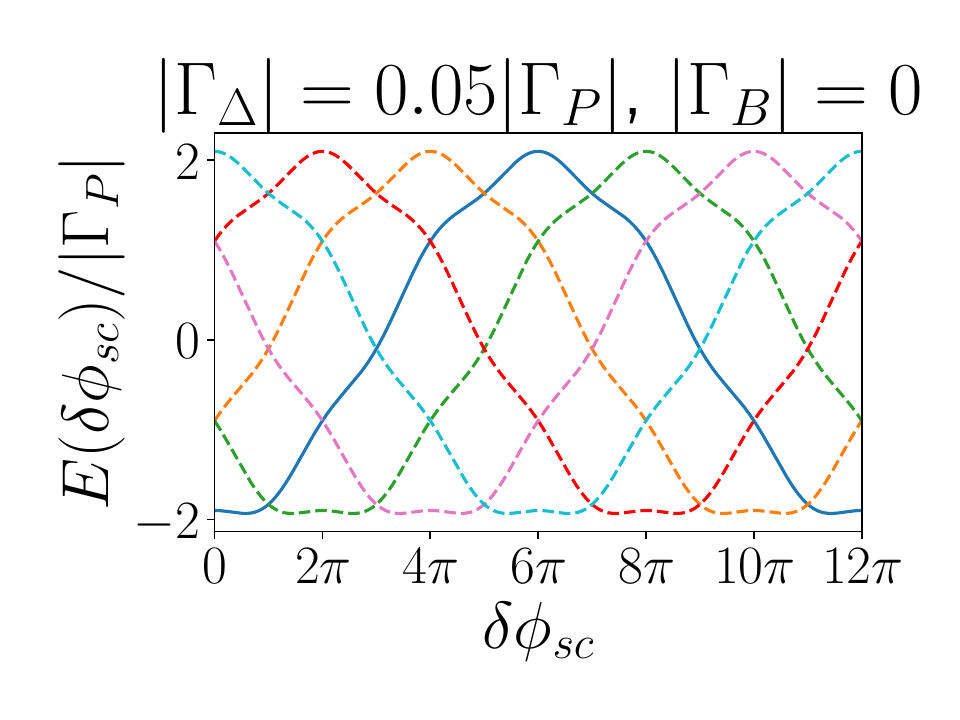}
         \captionsetup{justification=centering}
         \caption{}
         \label{fig:heff_PD}
    \end{subfigure}\hfill
    \begin{subfigure}{.5\linewidth}
         \centering
         \hspace*{-0.45cm}\includegraphics[width=\linewidth]{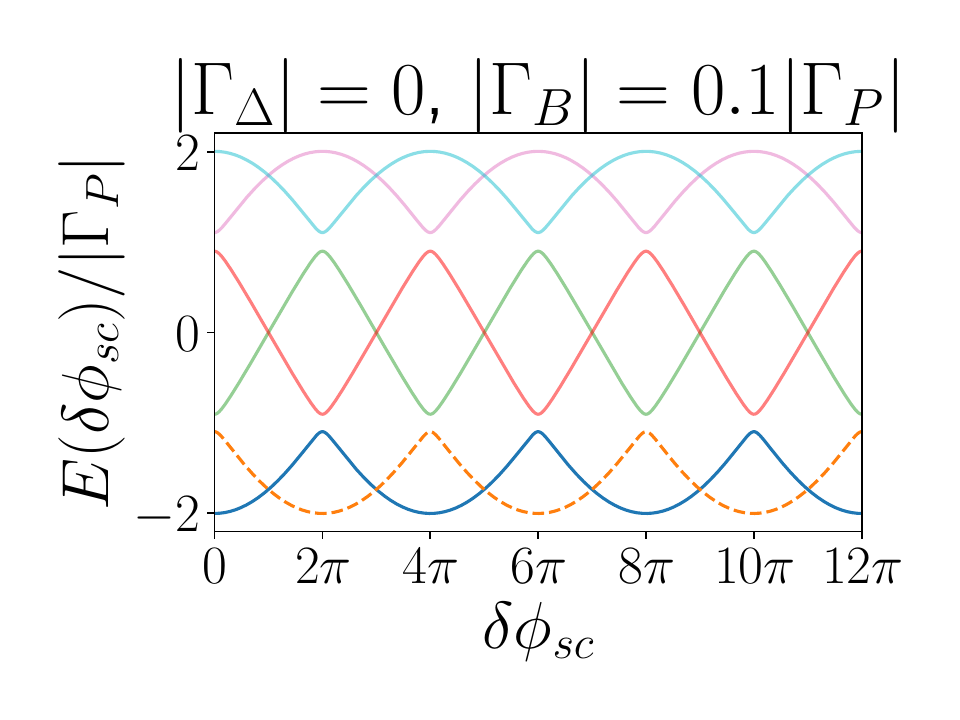}
         \captionsetup{justification=centering}
         \caption{}
         \label{fig:heff_PB}
    \end{subfigure}\hfill
    \begin{subfigure}{.5\linewidth}
         \centering
         \hspace*{-0.45cm}\includegraphics[width=\linewidth]{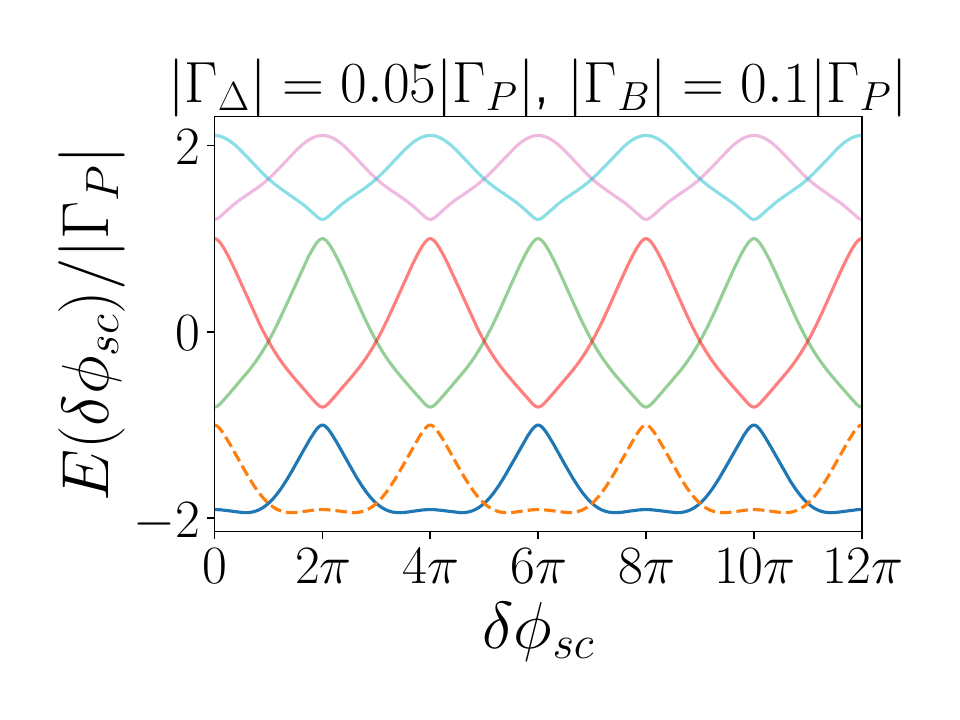}
         \captionsetup{justification=centering}
         \caption{}
         \label{fig:heff_PBD}
    \end{subfigure}
\caption{Spectra of tunneling effective Hamiltonian as functions of the Josephson phase $\de\phi_{sc}$ with (a) only parafermion tunneling, (b) parafermion and Cooper pair tunneling, (c) parafermion tunneling and backscattering and (d) backscattering, parafermion, and Cooper pair tunneling. Ground states at each $\de\phi_{sc}$ are indicated by thin lines. Dashed thin lines can be mapped to solid thin lines by a symmetry transformation. In (c) and (d), excited states are indicated with bold lines.}
\label{fig:heff}
\end{figure}
We can see the effect of each tunneling term from Eq. \eqref{eqn:heff} and Fig. \ref{fig:heff}. The Cooper pair tunneling term is proportional to identity; hence it only provides a $\de\phi_{sc}$-dependent shift to all states. With no backscattering, the eigenstates satisfy the boundary condition $\Psi_r(\de\phi_{sc}+2\pi)=\Psi_{r+1}(\de\phi_{sc})$ and the Hamiltonian is invariant under $\Z_{2m}$ transformation $n^{(1)}\ra n^{(1)}+1\mod2m$. The backscattering term explicitly breaks the $\Z_{2m}$ down to $\Z_2$, corresponding to $n^{(1)}\ra n^{(1)}+m \mod 2m$. The eigenstates now satisfy a different boundary condition, $\Psi_r(\de\phi_{sc}+2\pi)=\Psi_{r+m}(\de\phi_{sc})$. 

These effects demonstrate that the $\Z_m$ part of the symmetry is inherently different from the $\Z_2$ fermion parity. The unbroken $\Z_2$ represents the topologically protected fermion parity and can only be broken by nonlocal terms like tunneling of parafermion across the FTSC, $\al_{-L}^\da\al_0$, whereas the $\Z_m$ symmetry can be broken by local tunneling terms like tunneling of a quasiparticle and a quasihole $\psi_{R,qp}^{(1)\da}\psi_{L,qp}^{(1)}\psi_{R,qp}^{(2)}\psi_{L,qp}^{(2)\da}+h.c.\sim \cos[2(\vp_0^{(1)}-\vp_0^{(2)})]$. This suggests that in systems with backscattering, one cannot distinguish PZM from MZM tunneling from the periodicity of energy-phase relation since both of them have $\Z_2$ symmetry.

\section{Tunneling currents}
\label{sec:current}
\begin{figure}
    \begin{subfigure}{.5\linewidth}
         \centering
         \hspace*{-0.55cm}\includegraphics[width=\linewidth]{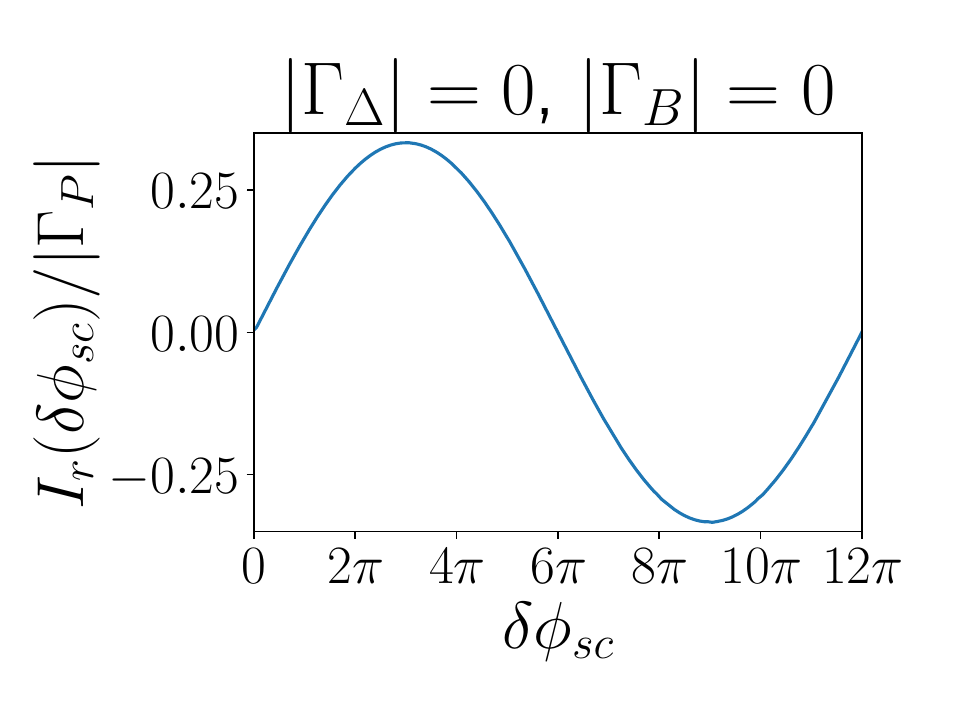}
         \captionsetup{justification=centering}
         \caption{}
         \label{fig:I_P}
    \end{subfigure}\hfill
    \begin{subfigure}{.5\linewidth}
         \centering
         \hspace*{-0.55cm}\includegraphics[width=\linewidth]{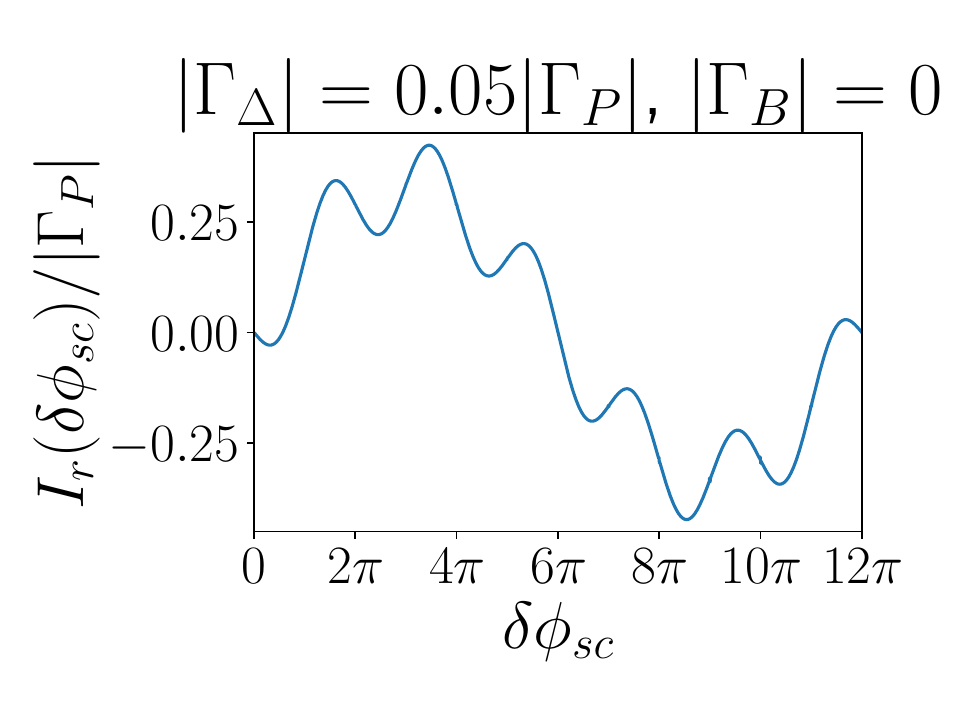}
         \captionsetup{justification=centering}
         \caption{}
         \label{fig:I_PD}
    \end{subfigure}\hfill\\
    \begin{subfigure}{.5\linewidth}
         \centering
         \hspace*{-0.55cm}\includegraphics[width=\linewidth]{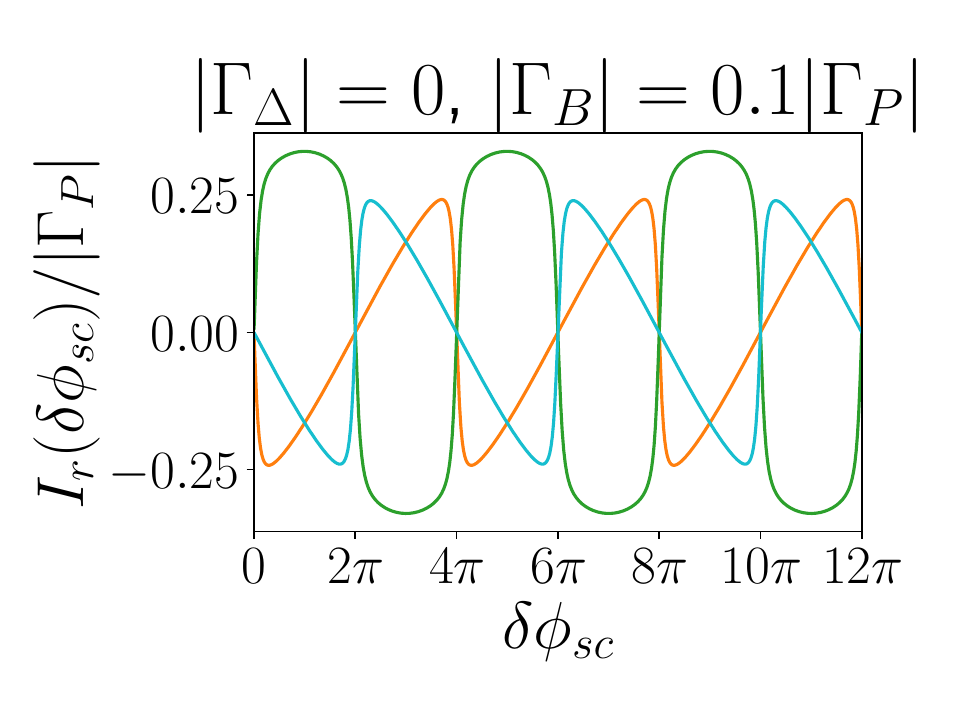}
         \captionsetup{justification=centering}
         \caption{}
         \label{fig:I_PB}
    \end{subfigure}\hfill
    \begin{subfigure}{.5\linewidth}
         \centering
         \hspace*{-0.55cm}\includegraphics[width=\linewidth]{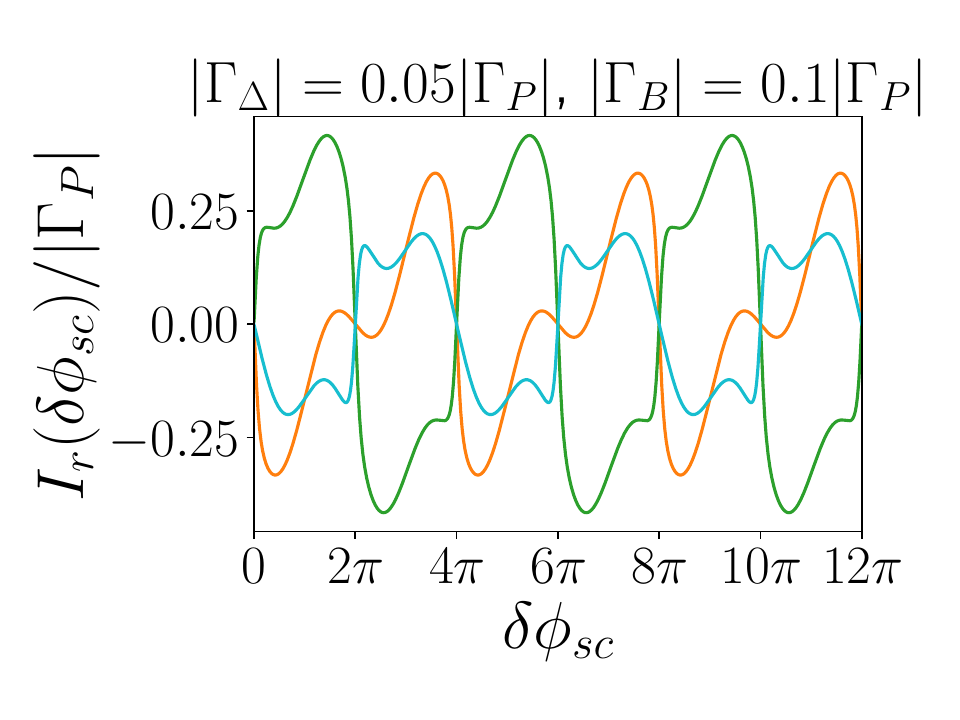}
         \captionsetup{justification=centering}
         \caption{}
         \label{fig:I_PBD}
    \end{subfigure}
\caption{Single-channel current-phase relations $I_r(\de\phi_{sc})$ with $\nu=1/3$ for (a) only parafermion tunneling, (b) parafermion and Cooper pair tunneling, (c) parafermion tunneling and backscattering, (d) backscattering, parafermion, and Cooper pair tunneling. Currents associated with additional parafermion states that are not shown are related to the currents in this figure by symmetry transformations of the eigenstates,  represented by $2\pi$ shifts in the Josephson phase. The currents have the same color as the corresponding energy states in Fig. \ref{fig:heff}.}
\label{fig:I}
\end{figure}

The difference between the $\Z_2$ and $\Z_m$ parts of the quasiparticle parity can also be seen in the current-phase relation where the Josephson phase is $4m\pi$-periodic without the backscattering term and $4\pi$-periodic with it, as shown in Fig.~\ref{fig:I}. From our effective Hamiltonian, the tunneling current operator is given by the commutator between tunneling Hamiltonian and the total number operator $\hat{N}^{(2)}=\pbr{\vp^{(2)}_{-L}-\vp^{(2)}_0}/\pi$,
\begin{align}
    \hat{I}(\de\phi_{sc})=&e\dv{\hat{N}^{(2)}}{t}=ie\comm{H_{\text{eff}}}{\hat{N}^{(2)}} \nonumber \\
    =&\frac{2e}{m}\abs{\Ga_P}\sin(\vt^{(1)}_0-\vt^{(2)}_0-\frac{\de\phi_{sc}}{2m})\label{eqn:Iop}\\
    &+4e\abs{\Ga_\De}\sin\sbr{2m\pbr{\vt^{(1)}_0-\vt^{(2)}_0}-\de\phi_{sc}}.\nonumber
\end{align}
As an operator, this definition is equivalent to $\hat{I}(\de\phi_{sc})=2e\dv{H_{\text{eff}}}{\de\phi_{sc}}$. The tunneling current for each eigenstate $\ket{\Psi_r}$ of Eq. \eqref{eqn:heff} is then given by,
\begin{align}
    I_r(\de\phi_{sc})=\frac{\ev{\hat{I}}{\Psi_r}}{\bk{\Psi_r}}=2e\dv{E_r}{\de\phi_{sc}}. \label{eqn:I_single}
\end{align}
We can see from Fig. \ref{fig:I}(c,d) that the backscattering term explicitly breaks the $\Z_m$ symmetry and results in a $4\pi$ periodicity in the Josephson current.

\begin{figure}
    \begin{subfigure}{.5\linewidth}
         \centering
         \hspace*{-0.7cm}\includegraphics[width=\linewidth]{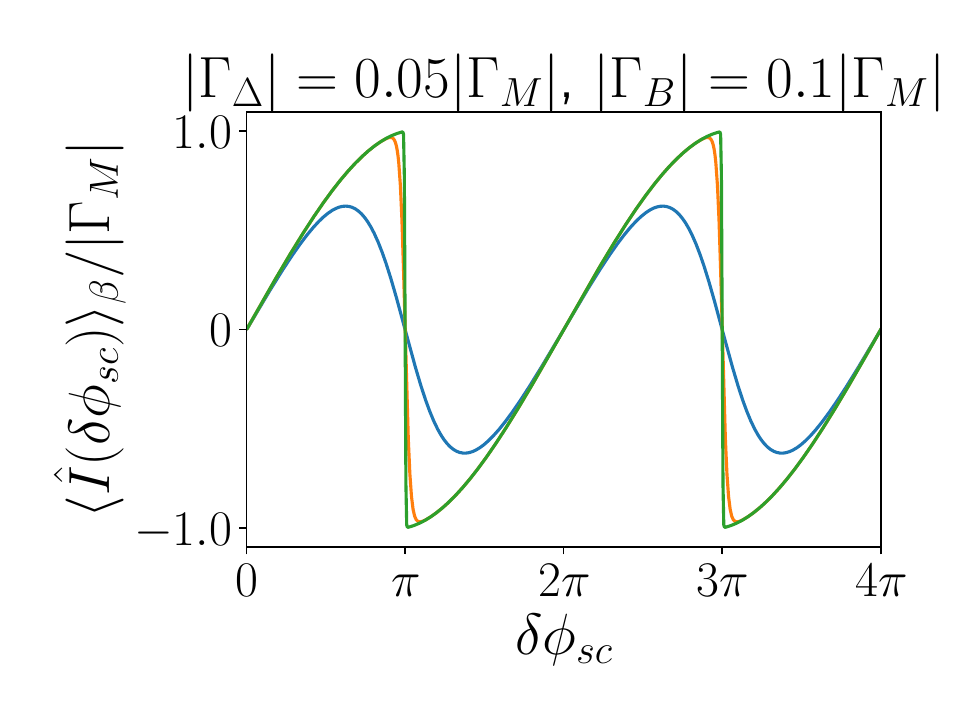}
         \captionsetup{justification=centering}
         \caption{}
         \label{fig:TI_beta_PBD1}
    \end{subfigure}\hfill
    \begin{subfigure}{.5\linewidth}
         \centering
         \hspace*{-0.65cm}\includegraphics[width=\linewidth]{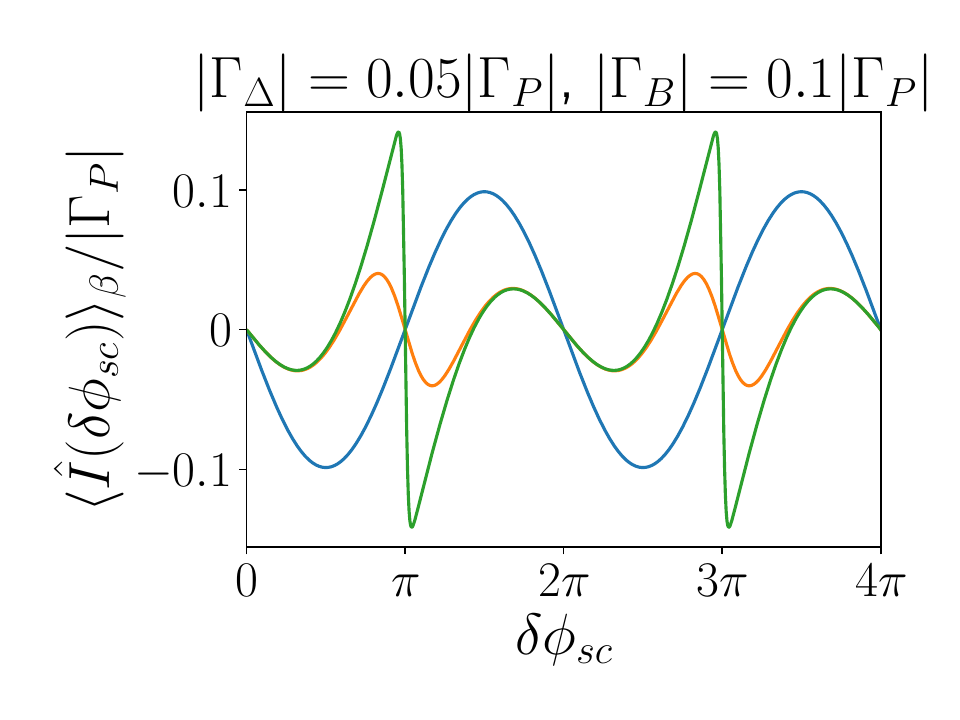}
         \captionsetup{justification=centering}
         \caption{}
         \label{fig:TI_beta_PBD}
    \end{subfigure}\hfill\\
    \begin{subfigure}{.5\linewidth}
         \centering
         \hspace*{-0.65cm}\includegraphics[width=\linewidth]{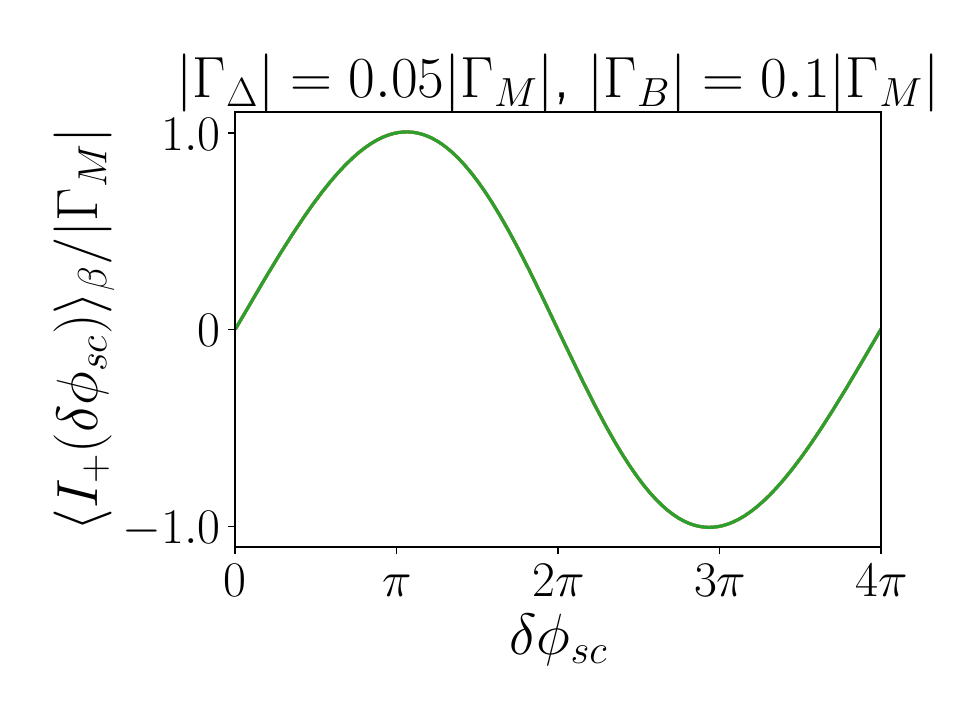}
         \captionsetup{justification=centering}
         \caption{}
         \label{fig:TI_beta_part_PBD1}
    \end{subfigure}\hfill
    \begin{subfigure}{.5\linewidth}
         \centering
         \hspace*{-0.65cm}\includegraphics[width=\linewidth]{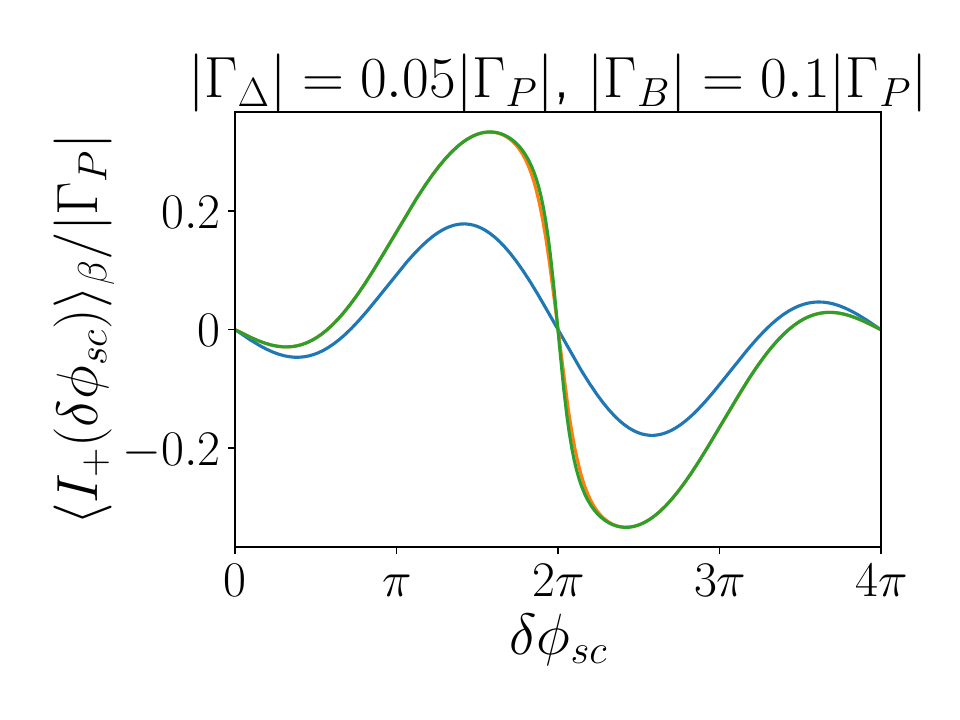}
         \captionsetup{justification=centering}
         \caption{}
         \label{fig:TI_beta_part_PBD}
    \end{subfigure}\hfill
\caption{Thermal average of the current-phase relation with all tunneling terms for different values of $\beta$ for (a) $\nu=1$ and (b) $\nu=1/3$. Projected thermal average of the current-phase relation with all tunneling terms for different values of $\beta$ for (c) $\nu=1$ and (d) $\nu=1/3$. $\beta=0.01$ (blue), 0.1 (orange), 1 (green) in units of $\abs{\Ga_M}^{-1}$ in (a,c) and $\abs{\Ga_P}^{-1}$ in (b,d).}
\label{fig:TI_beta}
\end{figure}

If no terms violate the total quasiparticle parity at the junction, e.g. terms proportional to $\al^{(1)\da}_L\al^{(1)}_0$, there will be no transitions between different channels of the tunneling current. However, at finite temperature, thermal excitations can break fermion parity where the thermal averaged current is,
\begin{align}
    \ev{\hat{I}(\de\phi_{sc})}_\be=\frac{\tr\pbr{e^{-\be H_{\text{eff}}}\hat{I}}}{\tr e^{-\be H_{\text{eff}}}}\label{eqn:I},
\end{align}
at inverse temperature $\be=1/T$. For $\nu=1$ with $\abs{\Ga_\De}=\abs{\Ga_B}=0$, Eq. \eqref{eqn:I} reduces to the known result in Ref. \cite{kwon-2004,fu-2009}. In the thermally-averaged current-phase relation Eq. \eqref{eqn:I}, each of the tunneling terms has different contributions. The thermally-averaged current-phase relation for parafermion and Majorana fermion tunneling are shown in Fig. \ref{fig:TI_beta}(a,b). Both parafermion and Majorana fermion terms exhibit a zig-zag pattern with a slope proportional to $\be$ whereas the Cooper pair tunneling term has a sine-wave contribution. There is a difference in the relative amplitude between Majorana fermion/parafermion tunneling and Cooper pair tunneling contribution in Eq. \eqref{eqn:Iop}. Backscattering does not contribute to the shape of the thermally-averaged current.  All thermally-averaged currents exhibit $2\pi$-periodicity due to the contributions from states with different quasiparticle parity.

The fractional Josephson effect can be detected at finite temperatures if one is capable of projecting into individual states. We can define the projected current using a projection operator $\hat{P}_r$ in the definition of trace in Eq. \eqref{eqn:I},
\begin{equation}
    \ev{I_r(\de\phi_{sc})}_\be=\frac{\tr(e^{-\be H_{\text{eff}}}\hat{P}_r\hat{I})}{\tr(e^{-\be H_{\text{eff}}}\hat{P}_r)}.
\end{equation}
These projection operators $\hat{P}_r$ are generalizations of the fermion parity projection operators $\hat{P}_{\pm}=[\id\pm (-1)^F]/2$, and $r$ depends on the parity symmetry of the system, i.e. it is $\pm$ if there is finite backscattering and ranges from 1 to $2m$ otherwise. The projection operators can be obtained by a linear combination of powers of the clock matrix $\si$ (see Appendix \ref{sec:projection} for details). In experiments, these projections can be realized by fixing the charge difference between the two FTSCs, represented by $n^{(1)}-n^{(2)}\mod 2m$ in Eq. \eqref{eqn:heff}. If there is no backscattering, the projected currents are equivalent to the single-channel currents in Eq. \eqref{eqn:I_single}, similar to the $\nu=1$ case in Fig. \ref{fig:TI_beta}(c). When there is backscattering, as in Fig. \ref{fig:TI_beta}(d), the parafermion tunneling adds a $4\pi$-periodic contribution to the current with the amplitude that scales as a power-law of inverse temperature $\be$. This behavior is unique to parafermion tunneling, and therefore, is a fingerprint of PZMs.

\section{Discussion}
\label{sec:discussion}
We have presented a model for FTSC consisting of two edge states from a $\nu=1/m$ FQH system proximity coupled by an s-wave superconductor and identified a PZM at one end of the FTSC. We have constructed an effective Hamiltonian and shown the roles played by parafermion tunneling, Cooper pair tunneling, and backscattering by identifying the symmetry of the ground states and Josephson periodicity in different tunneling currents. We showed that for a $\Z_{2m}$ FTSC, only the $\Z_2$ part of the symmetry is topologically protected while the $\Z_m$ part can be explicitly broken by local tunneling terms like tunneling of a quasiparticle and a quasihole. We proposed using a projected thermally-averaged current to detect parafermion tunneling, which has a $4\pi$-periodic fractional Josephson effect, in various superconducting qubit junctions \cite{calzona-2023,nielsen-2022,nielsen-2023}. We have also illustrated the different behaviors in junctions of FTSCs between the $\nu=1$ and the $\nu=1/3$ cases respectively.

Although the results we presented hold for general values of the tunneling amplitudes, the relative strength of the tunneling terms is important for experimental detection of the fractional Josephson effect, i.e., the parafermion tunneling term should be larger than the Cooper pair tunneling and the backscattering term. For a Josephson junction with gap distance $l$, the backscattering amplitude is expected to scale as $\Ga_B\sim e^{-l/\ell_0}$. The parafermion and Cooper pair tunneling amplitude are expected to scale as $\Ga_P,~\Ga_\De\sim e^{-l/\xi}$. As a realistic example, we can consider niobium nitride, which under a magnetic field of $\sim$6 T where $\nu=1/3$ has been shown to arise in graphene-based structures \cite{du-2009,dean-2011}, has a larger coherence length ($\sim$50 nm) than the magnetic length ($\sim$10 nm). Hence, we expect the backscattering to be much weaker than both Cooper pair and parafermion tunneling. We expect the Cooper pairing amplitude to be much smaller than the parafermion tunneling because the tunneling of multiple charges is suppressed through an FQH background. The effects described in this letter should therefore be observable in experiments. Additional screening layers may also be beneficial for entering the regime where parafermion tunneling dominates.

In this letter, we assumed an absence of disorder in the FQH background. For FQH background with disorder, the result would depend on details of the platform, and edge reconstruction would lead to a discrepancy between theory and experiment. For spin-unpolarized FQH background, there are additional effects from spin-charge separation but such a system is difficult to realize in a graphene-based device due to its untunable g-factor. In such a platform the spin up and down branches of the edge states are spatially separated due to the sizable Zeeman interaction, unlike in GaAs heterostructures where the g-factor can be tuned.

Our results can be generalized to FTSCs with FQH backgrounds at different filling fractions. For clean, spin-polarized FQH systems in the Jain sequence \cite{jain-1989}, the proximity effect on the spatially outermost edge is stronger than that on the inner edges \cite{gul-2022}. We expect to see only the behavior corresponding to the outermost edge. For example, with $\nu=2/5$ FQH background, the thermal-averaged current-phase relations should be similar to $\nu=1/3$ in Fig. \ref{fig:TI_beta} (b,d).

We thank Erez Berg, Paul Fendley, Netanel Lindner, Yi-Zhuang You, and Lucas Wagner for insightful discussions. This work was supported in part by the US National Science Foundation (NSF) through the grant NSF DMR-2225920 at the University of Illinois (JC and EF) and through the NSF Quantum Leap Challenge Institute for Hybrid Quantum Architectures and Networks, NSF OMA-2016136 (AK).
\appendix
\section{Review of PZM in the N-state quantum clock model}
\label{sec:clock}
Here we review the $N$-state clock model \cite{fradkin-1980} which can host PZMs with an open boundary condition. The $N$-state clock model is a $\Z_N$ generalization of the Ising spin chain defined on a one-dimensional lattice with $L$ sites,
\begin{align}
	H_{\text{clock}}=-\sum_{j=1}^{L-1} \pbr{\si^\da_j \si_{j+1}+\hc}-\la\sum_{j=1}^L \pbr{\tau^\da_j+\tau_j},\label{eqn:clock}
\end{align}
where $\la>0$ is a coupling constant; $\si$ and $\tau$ are the ``clock" and ``shift" operators \cite{fradkin-1980}. These operators satisfy,
\begin{equation}
    \begin{split}
        \si_j^N=\tau_j^N=\id,\quad \si_j^\da&=\si_j^{N-1},\quad \tau_j^\da=\tau_j^{N-1},\\
    \si_j\tau_j&=\om\tau_j\si_j,
    \end{split}
\end{equation}
where $\om\equiv e^{2\pi i/N}$. In the representation where $\si_j$ is diagonalized, $\si_j$ and $\tau_j$ have the following matrix representation,
\begin{align}
	\si_j=\mqty(\dmat{1,\om,\om^2,\ddots,\om^{N-1}}),~\tau_j=\mqty(0&0&0&\cdots &0&1\\1&0&0&\cdots &0&0\\0&1&0&\cdots &0&0\\\vdots &&&&\vdots &\vdots\\0&0&0&\cdots &1&0)\label{eqn:clock_mat}.
\end{align}
With the Fradkin-Kadanoff transformation \cite{fradkin-1980}, the clock model Hamiltonian can be written in terms of parafermion operators $\al_k$,
\begin{align}
	\al_{2j-1}=\si_j\prod_{i<j}\tau_j,\quad \al_{2j}=-e^{i\pi /N}\tau_i\si_j\prod_{i<j}\tau_i.
\end{align}
These parafermion operators satisfy the parafermion algebra,
\begin{equation}
    \begin{split}
	\al_{k}^N&=\id,\quad\al_k^\da=\al_k^{N-1},\\
 \al_k\al_{k'}&=e^{i\frac{2\pi}{N}\sgn(k-k')}\al_{k'}\al_k,\label{eqn:para}
    \end{split}
\end{equation}
and the clock model Hamiltonian in the parafermion basis writes,
\begin{equation}
    \begin{split}
        H_{\text{clock}}=&\sum_{j=1}^{L-1}\pbr{e^{-i\frac{\pi}{N}}\al^\da_{2j}\al_{2j+1}+\hc}\\
 &+\la\sum_{j=1}^L\pbr{e^{i\frac{\pi}{N}}\al^\da_{2j-1}\al_{2j}+\hc}.
    \end{split}
\end{equation}
The clock model reduces to the Kitaev chain \cite{kitaev-2001} and the parafermion operators become Majorana fermion operators when $N=2$. Similar to the Kitaev chain, when $\la\ll1$, the first term dominates and couples the parafermions on the same site. This leads to a unique ground state and a gapped spectrum, i.e. a trivial phase. For $\la\gg 1$, the second term dominates and the chain now supports two PZMs $\al_1$ and $\al_{2L}$ at the ends. There are $N$ distinct eigenvalues for $\al^\da_{2L}\al_1$ (that does not affect energy); hence, there are $N$ degenerate ground states in this topological phase.

A parafermion zero mode operator commutes with the Hamiltonian, and has a non-trivial commutation relation with a charge operator $P\equiv\prod_{j}\tau_j^\da$, which is a $\Z_N$ generalization of the fermion parity $(-1)^F$ \cite{fendley-2012},
\begin{align}
	\comm{H_{\text{clock}}}{\al}=0,~P\al=\om\al P.
\end{align}
$P$ is the symmetry generator of the $\Z_N$ symmetry $\si_j\ra\om\si_j$ for all $j$ representing ``turning the clocks for all sites" and it satisfies $P^N=\id$. 

\section{Bosonization and Klein Factors}
\label{sec:bosonization}
The electron and quasiparticle operators in FTSC $i$ on the top/bottom edge (R/L) can be written as bosonic fields using the standard (1+1)-D bosonization procedure,
\begin{equation}
    \begin{split}
        &\psi_{R}\sim F^{(i)}e^{im\phi^{(i)}_{R}},\quad\psi_{L}\sim F^{(i)\da}e^{-im\phi^{(i)}_L},\\
        &\psi_{qp,R}\sim F_{qp}^{(i)}e^{i\phi^{(i)}_{R}},\quad\psi_{qp,L}\sim F_{qp}^{(i)\da}e^{-i\phi^{(i)}_L},
    \end{split}
\end{equation}
where $F^{(i)}$ and $F^{(i)}_{qp}$ are Klein factors for electrons and quasiparticles to keep operators' commutation relation on different FTSCs consistent. The bosonic fields satisfy the commutation relations, 
\begin{equation}
    \begin{split}
        \comm{\phi_{R/L}(x)}{\phi_{R/L}(x')}&=\pm i \frac{\pi}{m}\sgn(x-x'),\\
        \comm{\phi_{L}(x)}{\phi_{R}(x')}&=i\frac{\pi}{m}, 
    \end{split}
\end{equation}
and the Klein factors satisfy,
\begin{equation}
    \begin{split}
        \acomm{F^{(i)\da}}{F^{(j)}}&=2\de_{ij},\\
        \acomm{F^{(i)\da}}{F^{(j)\da}}&=\acomm{F^{(i)}}{F^{(j)}}=0,\\
    F^{(i)}_{qp}F^{(j)}_{qp}&=e^{i\frac{\pi}{m}\sgn(i-j)}F^{(j)}_{qp}F^{(i)}_{qp},\\
    F^{(i)}_{qp}F^{(j)\da}_{qp}&=e^{i\frac{\pi}{m}}F^{(j)\da}_{qp}F^{(i)}_{qp},
    \end{split}
\end{equation}
The quasiparticle Klein factors and the electron Kelin factors are related by $\left(F_{qp}^{(i)}\right)^m=F^{(i)}$. Since all of the Klein factors are unitary, $F^{(i)\da}F^{(i)}=F^{(i)}F^{(i)\da}=F_{qp}^{(i)\da}F_{qp}^{(i)}=F_{qp}^{(i)}F_{qp}^{(i)\da}=1$, they will not affect any physical observables in the main text. However, they play an important role in parafermion braiding.
\section{Kosterlitz renormalization group for sine-Gordon theory}
\label{sec:RG}
In this section, we aim to present a detailed calculation of the RG flow. The Lagrangian of the FTSC is,
\begin{equation}
    \L=\frac{m}{2\pi K}\pbr{\pa_\mu\vt}^2+\frac{\De}{\ell_0^2}\cos(2m\vt).
\end{equation}
Rescaling the $\vt\ra\vt'=\sqrt{\frac{m}{\pi K}}\vt$, we obtain the sine-Gordon Lagrangian,
\begin{equation}
    \L=\frac{1}{2}\pbr{\pa_\mu\vt}^2+\frac{\De}{\ell_0^2}\cos(\be\vt),
\end{equation}
with $\be=2\sqrt{m\pi K}$. In the vicinity of the critical point, the proximity gap $\De$ and the distance from the Luttinger parameter to the critical point $x\equiv 2-mK$ follow Kosterlitz RG equations \cite{kosterlitz-1974,Amit-1980},
\begin{figure}
    \centering
    \includegraphics[scale=0.45]{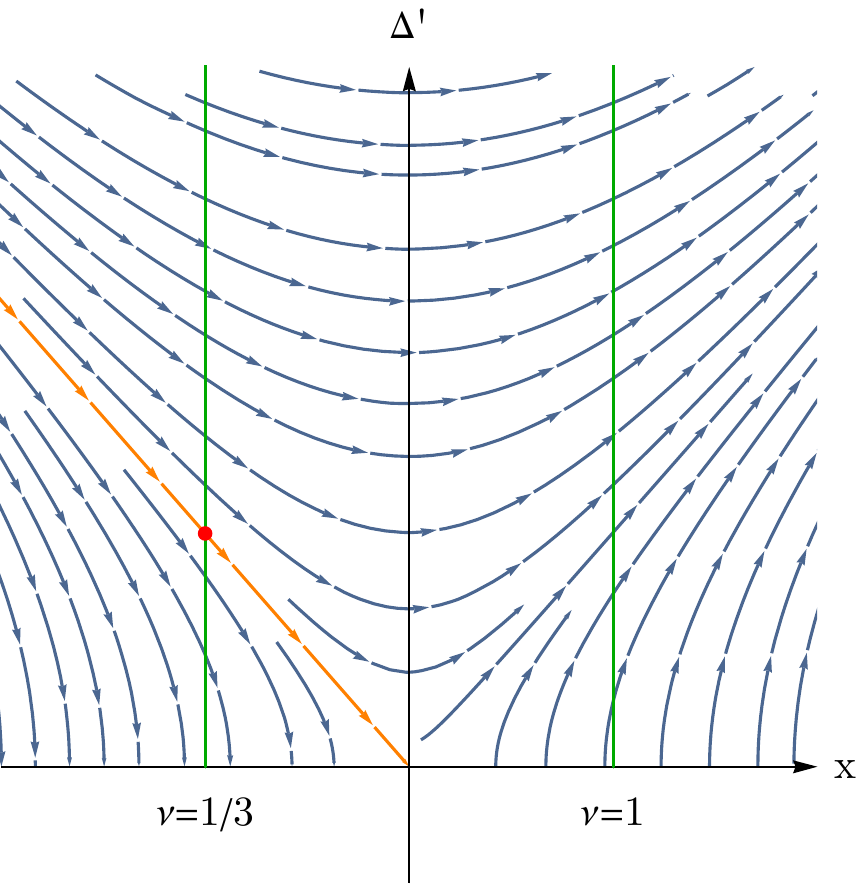}
    \caption{RG flow for $x$ and $\Delta'$. The orange line $\De_c'=-x$ is the locus of the phase transition between the proximity-induced superconducting phase ($\Delta'>-x$ and the non-superconducting phase. For a given value of $K>K_c$ there is a given (negative) value of $x$. At fixed $x$ there will be a phase transition to the superconducting state given in Eq. \eqref{eq:critical-Delta}. The green lines represent values of $x$ without any interaction, which means that for $\nu=1$ there is always a proximity effect whereas for $\nu=1/3$ a critical $\De_c$ is needed for the system to be superconducting. This estimate is accurate only for $K$ close to $K_c=2/m$.}
    \label{fig:RG}
\end{figure}
\begin{align}
    \dv{\De}{l}=x\De+...,~\quad \dv{x}{l}=128m^2\pi^5\De^2+...
\end{align}
A rescaling of $\De\ra\De'=8\sqrt{2}m\pi^{5/2}\De$ could eliminate the coefficient in RG flow of $x$. With a repulsive Coulomb interaction $U>0$ and a Luttinger parameter $K$, the critical proximity gap $\De_c$ needed for the system to be in the superconducting phase is,
\begin{equation}
    \De_c\simeq \frac{mK-2}{8\sqrt{2}m\pi^{5/2}}
    \label{eq:critical-Delta}
\end{equation}
The actual proximity gap is proportional to $\Delta$ multiplied by a transparency factor determined by how well the superconductor couples to the edge states of the quantum Hall fluid.  

\section{Projection Operators}
\label{sec:projection}
Here we present the projection operators $\hat{P}_r$ in terms of clock model matrix $\si$ and $\om\equiv e^{i\pi/3}$,
\begin{equation}
    \begin{split}
        \hat{P}_1&=\frac{1}{6}\pbr{\id-\om^2\si-\om\si^2-\si^3+\om^2\si^4+\om\si^5},\\
        \hat{P}_2&=\frac{1}{6}\pbr{\id-\om\si+\om^2\si^2+\si^3-\om\si^4+\om^2\si^5},\\
        \hat{P}_3&=\frac{1}{6}\pbr{\id-\si+\si^2-\si^3+\si^4-\si^5},\\
        \hat{P}_4&=\frac{1}{6}\pbr{\id+\om^2\si-\om\si^2+\si^3+\om^2\si^4-\om\si^5},\\
        \hat{P}_5&=\frac{1}{6}\pbr{\id+\om\si+\om^2\si^2-\si^3-\om\si^4-\om^2\si^5},\\
        \hat{P}_6&=\frac{1}{6}\pbr{\id+\si+\si^2+\si^3+\si^4+\si^5},
    \end{split}
\end{equation}
where $r$ represents the eigenvalue of $n^{(1)}$, which fixes the charge difference between two FTSCs $n^{(1)}-n^{(2)}$ for a fixed total quasiparticle parity $n^{(1)}+n^{(2)}$. For systems with finite backscattering, the projection operators become $\hat{P}_\pm=\frac{1}{2}\pbr{\id\pm\si^3}$.
\bibliography{parafermion}
\end{document}